\documentclass[twocolumn,twoside,slac_two]{revtex4}
\usepackage{graphicx}
\usepackage{fancyhdr}
\usepackage{wasysym}
\usepackage{threeparttable}
\newcommand{\gammaray}{\mbox{$\gamma$-ray}}

\newcommand{\fluxunits}{cm$^{-2}$\,s$^{-1}$}

\def\flux{\textrm{TeV}^{-1}\textrm{cm}^{-2}\textrm{s}^{-1}}
\def\hess{H.E.S.S.}

\def\psrb{PSR B1259-63/LS 2883}
\def\pulsar{PSR B1259-63}

\newcommand{\beq}{\begin{equation}}
\newcommand{\enq}{\end{equation}}

\begin{document}

\title{VHE Emission from \psrb\ around 2010/2011 Periastron Passage observed with \hess}

\author{I. Sushch}
\affiliation{Institut f\"ur Physik, Humboldt-Universit\"at zu Berlin, Newtonstr. 15, D 12489 Berlin, Germany}
\affiliation{Astronomical Observatory of Ivan Franko National University of L'viv, vul. Kyryla i Methodia, 8, L'viv 79005, Ukraine}
\affiliation{Centre for Space Research, North-West University, Potchefstroom, South Africa}

\author{M. de Naurois}
\affiliation{Laboratoire Leprince-Ringuet, Ecole Polytechnique, CNRS/IN2P3, F-91128 Palaiseau, France}

\author{U. Schwanke, G. Spengler}
\affiliation{Institut f\"ur Physik, Humboldt-Universit\"at zu Berlin, Newtonstr. 15, D 12489 Berlin, Germany}
\author{P. Bordas}
\affiliation{Institut f\"ur Astronomie und Astrophysik, Universit\"at T\"ubingen, Sand 1, D 72076 T\"ubingen, Germany}
\author{on behalf of the H.E.S.S. Collaboration}

\vspace{-10pt}
\begin{abstract}
PSR B1259-63/LS 2883 is a binary system consisting of a 48 ms pulsar orbitting around a Be star with an orbital period of  $\sim 3.4$ years. The system
was detected at very high energies (VHE;  $E>100$ GeV) by the High Energy Stereoscopic System (H.E.S.S.) during its periastron passages in 2004 and 2007.
Here we present new H.E.S.S. observations corresponding to its last periastron passage, which occurred on December 15th 2010. These new observations
partially overlap with the beginning of a spectacular gamma-ray flare reported by the {\it{Fermi}}-LAT. The H.E.S.S. observations show  both flux and
spectral properties similar to those reported in  previous periastron passages, without any signature of the emission enhancement seen at GeV energies. A
careful statistical study based on the Fermi and H.E.S.S. lightcurves leads to the conclusion that the GeV and TeV emission during the flare have a
different physical origin. This conlusion, in turn, allows to use 
{\it{Fermi}}-LAT measurements of the GeV flux as upper limits for the modeling 
of the VHE emission.

\end{abstract}

\maketitle

\thispagestyle{fancy}


\section{Introduction}
The class of very high energy (VHE; $E>100$ GeV) $\gamma$-ray binaries 
comprises only a handful of known objects in our Galaxy: 
LS 5039 \citep{LS_5039}, LS I +61 303 \citep{LS_61303}, 
\psrb\ \citep{psrb1259_hess05} and HESS J0632+057 \citep{hess_j0632}, 
the first binary primarily discovered at TeV energies. 
%% This class can be extended by two more candidates: Cygnus X-1 \citep{cygnusX}, 
%% a stellar-mass black hole binary detected at VHEs at the 
%% 4.1 $\sigma$ significance level, and HESS J1018-589 \citep{hess_j1018}, 
%% which shows a point-like component spatially coincident 
%% with the GeV binary 1FGL J1018.6-5856 
%% discovered by the Fermi LAT collaboration \citep{Fermi_catalog1}.  
Only for the source presented in this paper, \psrb, the compact companion 
is clearly identified as a pulsar, making it a unique object for the 
study of the interaction between pulsar and stellar winds 
and the emission mechanisms in such systems. 

It consists of a rapidly rotating pulsar with a 
spin period of $\simeq 48$ ms and a spin-down luminosity of 
$\simeq 8\times 10^{35}$ erg/s in a highly eccentric 
($e = 0.87$) orbit around a 
massive Be star \citep{johnston1, johnston2}. 
The pulsar moves around the companion with a period $P_{\mathrm{orb}} = 3.4$ 
years (1237 days). The luminosity of the companion star 
LS 2883 is $L_{\ast}=2.3\times10^{38}$\,erg\,s$^{-1}$ \citep{negueruela}.
Because of its fast rotation, the star is oblate with an equatorial 
radius of $R_{\mathrm{eq}} = 9.7\,R_{\astrosun} $ and a polar radius 
of $R_{\mathrm{pole}} = 8.1\,R_{\astrosun}$. This leads to a strong 
gradient of the surface temperature from $T_{\mathrm{eq}} \approx 27,500 K$ 
at the equator to $T_{\mathrm{pole}} \approx 34,000 K$ at the poles. 
%% The mass function of the system suggests a mass of the star 
%% $M_{\ast} \approx 30 M_{\astrosun}$ and an orbital inclination angle 
%% $i_{\mathrm{orb}} \approx25^{\circ}$ for the smallest neutron star mass of 
%% $1.4 M_{\astrosun}$. 
The optical observations also 
suggest that the system is located at the same distance as the 
star association Cen OB1 at $d = (2.3\pm 0.4)$ kpc \citep{negueruela}. 
The companion Be star features an equatorial disk that 
is believed to be inclined with respect to the pulsar's orbital plane 
\citep{johnston1, melatos, negueruela} in a way that the pulsar crosses 
the disk twice in each orbit just before ($\sim 20$\,days) and just 
after ($\sim 20$\,days) the periastron. 

Since its discovery in 1992, \psrb\ is constantly monitored by various 
instruments at all energy bands. 
%% The source shows broadband emission 
%% and is visible from radio wavelengths up to the VHE regime. 
The properties of 
the radio emission differ depending on the distance between 
the pulsar and the star. Radio observations \citep{johnston99, connors2002, 
johnston2005} show that 
when the pulsar is far from the periastron, the observed radio 
emission consists only of the pulsed 
component, whose intensity is almost independent 
of the orbital position. But closer to 
the periastron, starting at about $t_{\mathrm{p}}-100$ d, where 
$t_{\mathrm{p}}$ is the time of periastron, the intensity starts 
to decrease up to the complete disappearance approximately at $t_{\mathrm{p}}-20$ d. 
This is followed by an eclipse of the pulsed emission for about 
35-40 days when the pulsar is behind the disk. In contrast, a transient 
unpulsed component appears and sharply rises to a level more than 
ten times higher than the flux density 
of the pulsed emission far from the periastron. 
%% The unpulsed component is believed to come from 
%% synchrotron radiation generated in the 
%% shocked wind zone between the relativistic pulsar wind and 
%% the stellar disk outflow. 
After the disk crossing the unpulsed emission shows a 
slight decrease with another increase around $t_{\mathrm{p}} + 20$ d 
at the second crossing of the disk. %% Radio observations 
%% around the 2007 periastron passage showed extended unpulsed emission 
%% with a total projected extent of $\sim120$ AU and the peak of the emission 
%% clearly displaced from the binary system orbit \citep{moldon2011}. This indicates that 
%% a flow of synchrotron-emitting particles, which can travel far away 
%% from the system, can be produced in \psrb. 

\psrb\ is very well covered by X-ray observations carried out with various 
instruments such as \emph{ROSAT} \citep{cominsky94}, 
\emph{ASCA} \citep{kaspi95, hirayama99}, 
\emph{INTEGRAL} \citep{shaw2004}, \emph{Suzaku}, \emph{Swift}, \emph{Chandra} 
\citep{chernyakova2009}, and \emph{XMM-Newton} 
\citep{chernyakova2006, chernyakova2009}. 
%% The periastron passage in 2007 was monitored at the same time 
%% by \emph{Suzaku}, \emph{Swift}, \emph{XMM-Newton}, and \emph{Chandra} \citep{chernyakova2009}. 
X-ray observations did not show any X-ray pulsed emission from the pulsar. 
Unpulsed non-thermal radiation from the source varies with orbital phase 
in both flux and spectral index. Similarly to radio measurements, 
the enhancement 
of the flux occurs shortly before and shortly after the periastron. 
Unambiguously, the enhancement of the non-thermal emission results 
from the interactions of the pulsar wind with the circumstellar disk 
close to the periastron passage. 

\psrb\ was observed by \hess\ around the periastron passages in 
2004 \citep{psrb1259_hess05} and 2007 \citep{psrb1259_hess09}, leading 
to a firm detection on both occasions. In 2004, \psrb\ was observed mostly 
after the periastron, in 2007 mostly before it. Therefore, 
the repetitive behaviour of the source, i.e. the recurrent appearance 
of the source near periastron at the same orbital phase, 
with the same flux level 
and spectral shape of the emission, was not precisely confirmed, 
since the observations covered different orbital phases. However, the similar 
dependence of the flux on the separation distance between 
the pulsar and the star 
for both periastron passages provides a strong indication 
for the repetitive behaviour \citep{kerschhaggl_2011}. 
\psrb\ was not detected in observations performed far 
from periastron in 2005 and 2006, 
which comprised $8.9$ h  and $7.5$ h of exposure, respectively.

Observations around the 2004 and 2007 periastron passages 
showed a variable behaviour of the source flux with time. 
A combined light curve of those two 
periastron passages indicates two 
asymmetrical peaks around periastron with a 
significant decrease of the flux at the periastron itself. 
Peaks of the TeV emission roughly coincide with the flux enhancement observed 
in other wavebands as well as with the eclipse of the pulsed radio emission, 
which indicates the position of the circumstellar disk. 
This coincidence suggests that the 
TeV emission from \psrb\ may be connected to the interaction 
of the pulsar with the disk.
\section{Multiwavelength Observations around the 2010 Periastron Passage}
H.E.S.S. observations around the most recent periastron 
passage which took place on 15th of December 2010 were 
performed as part of an extended multiwavelength (MWL) campaign 
including also radio, optical, X-ray and high energy 
(HE; $E>100$ MeV) observations. A joint MWL paper is 
in preparation.

The first results of the radio and X-ray monitoring of the source 
can be found in \citet{psrb1259_fermi}. The pulsed radio emission 
was monitored with Parkes telescope revealing an eclipse of the pulsed signal 
lasting from $t_{\mathrm{p}}-16$ d to $t_{\mathrm{p}}+$15 d. 
Radio emission from \pulsar\ at frequencies between 1.1 and 10 
GHz was observed using the ATCA array in the period from 
$t_{\mathrm{p}}-31$ d to $t_{\mathrm{p}}+55$ d. The detected unpulsed  emission 
around the periastron passage showed a behaviour similar to 
the one observed during previous observations. % \citep{psrb1259_radio_2004}.
The X-ray energy band was covered by three instruments: Swift, 
Suzaku and XMM-Newton. Observations confirmed the 1-10 keV lightcurve shape 
obtained in previous periastron observations, showing a rapid X-ray 
brightening starting at about $t_{\mathrm{p}}-25$ d with 
a subsequent decrease closer to periastron and a second increase 
of the X-ray flux after periastron \citep{psrb1259_fermi}.

Observations of the binary system \psrb\ at HEs 
were performed using the Large Area Telescope 
(LAT) on board of \emph{Fermi}. The data taken around the periastron 
passage were analysed by two independent working groups 
\citep{psrb1259_fermi, tam}, yielding similar results for 
the flaring period (see below), although there are some discrepancies 
related to the first detection period close 
to the periastron passage. Those differences do not affect however 
the conclusions drawn in this paper. 
% \footnote{Note, that apart from similiarities there are also some 
% discrepancies in analysis results of two working groups 
% which are related to the first detection period and may be 
% caused by the low statistics.}. 
The source was detected close to periastron 
with a very low photon flux above 100 MeV of about 
$(1-2)\times10^{-7}$ \fluxunits. After the initial 
detection, the flux decreased and the source was below 
the detection threshold 
%undetected 
% for about a month from mid-December 
% to mid-January. 
until 14th of January, $t_{\mathrm{p}}+30$ d, when a spectacular flare was detected 
which lasted for about 7 weeks with 
an average flux $\sim10$ times higher than the flux detected 
close to the periastron \citep{psrb1259_fermi}. 
The highest day-averaged flux during 
the flare almost reached the estimate of the spin-down luminosity 
of the pulsar. 
%% , which indicates a close to 100\% efficiency of the 
%% conversion of the pulsar's rotational energy into $\gamma$-rays.
The HE emission around the periastron 
as a function of time significantly differs from 
the two-peak lightcurves observed in other wavebands. 
The flare is not coincident with the post-periastron 
peak in radio, X-rays, and VHE $\gamma$-rays. It is also 
much brighter in comparison to the first detection 
of GeV emission close to the periastron 
passage \citep{psrb1259_fermi, tam}. 
\section{H.E.S.S. Observations around 2010/2011 Periastron Passage}
\subsection{H.E.S.S. Observations Results}
\begin{figure}
\centering
%\vspace{-10pt}
\resizebox{\hsize}{!}{\includegraphics{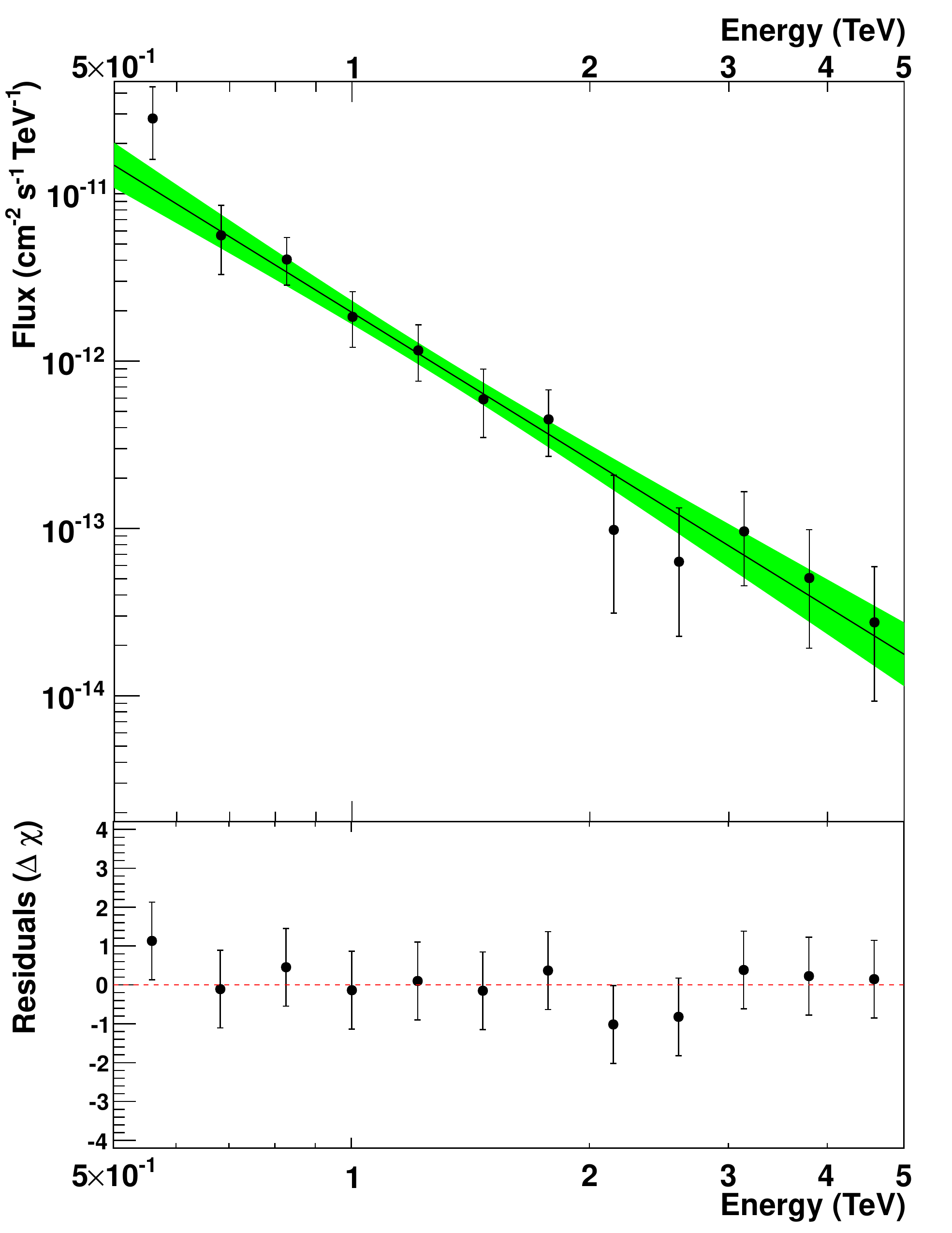}}
\vspace{-10pt}
\caption{Overall differential energy spectrum of the VHE \gammaray\ emission from \psrb\ for the whole observation period from $9^{\mathrm{th}}$ to $16^{\mathrm{th}}$ of January 2011. The solid line denotes the spectral fit with a simple power law. 
The green band represents the 1 $\sigma$ confidence interval. Points are derived for the minimum significance of $1.5 \sigma$ per bin. Points' error bars represent $1 \sigma$ errors. The figure is taken from \citet{sushch_psrb}.} 
  \vspace{-10pt}
  \label{spectrum_psrb1259}
\end{figure}

%\subsection{Data set and analysis techniques}
%\label{analysis}
\psrb\ observations were taken in five nights, 
namely January 9/10, 10/11, 13/14, 14/15 
and 15/16, resulting in 6.2 h of data \citep{sushch_psrb}. 
The source was detected at an 11.5 $\sigma$ level. 
A spectral analysis of the detected excess events 
shows that the differential energy spectrum of photons 
is consistent with a simple power law $\mathrm{d}N/\mathrm{d}E = 
N_{0} \left( E/1 \mathrm{TeV} \right)^{-\Gamma}$
with a flux normalisation at 1 TeV of $N_{0} = (1.95\pm0.32_{\mathrm{stat}}\pm 0.39_{\mathrm{syst}})\times10^{-12}\, \flux$ 
and a spectral index $\Gamma = 2.92\pm0.30_{\mathrm{stat}} \pm 0.20_{\mathrm{syst}}$ 
(see Fig. \ref{spectrum_psrb1259}) 
with a fit probability of 0.64. The integral flux above 1 TeV averaged over the 
whole observation period is $F(E>1\mathrm{TeV})=(1.01\pm0.18_{\mathrm{stat}} 
\pm 0.20_{\mathrm{sys}})\times10^{-12}$ \fluxunits\ \citep{sushch_psrb}.

\begin{figure}
  \centering
  \resizebox{\hsize}{!}{\includegraphics[width=\textwidth]{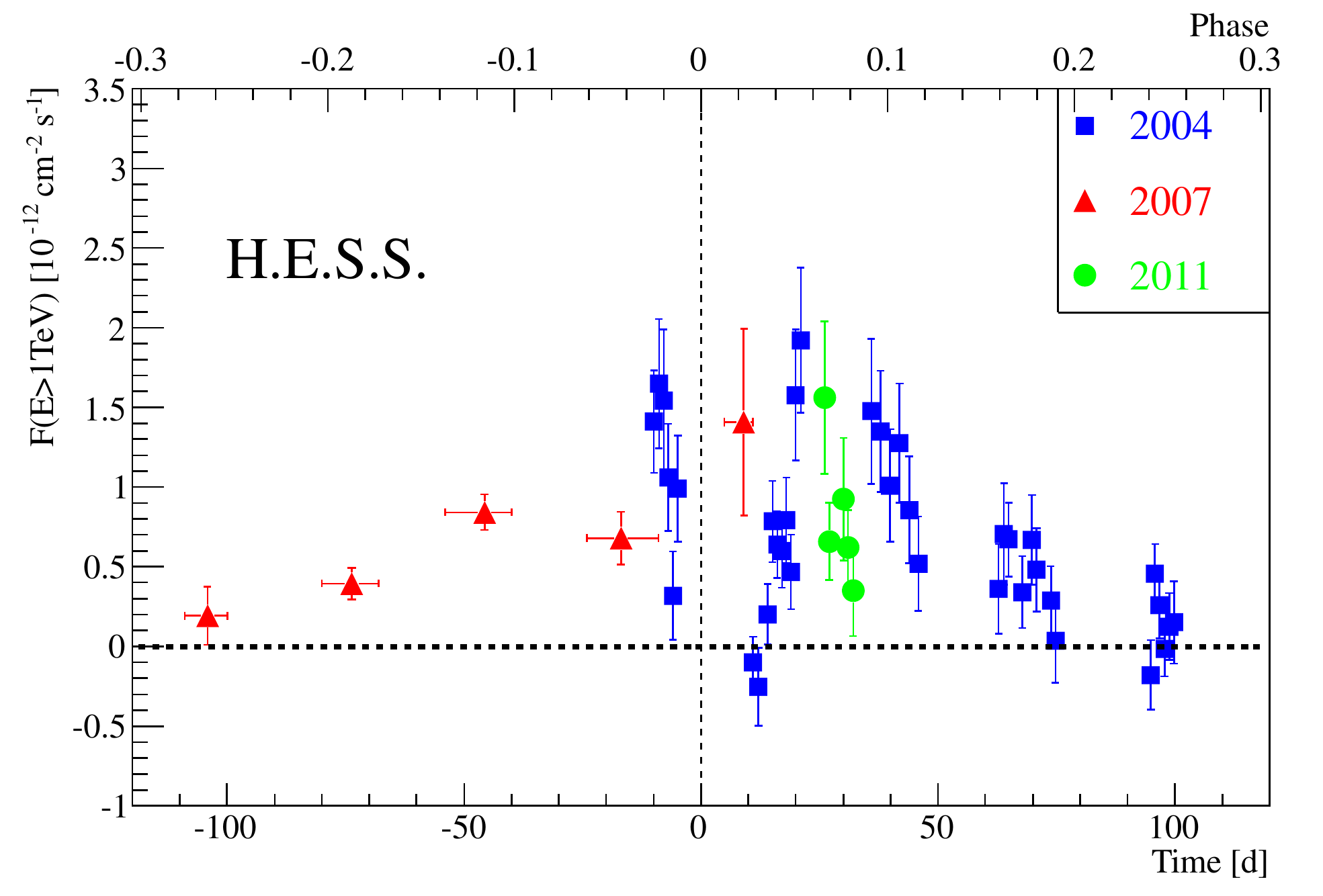}}
   \vspace{-10pt}
  \caption{Integrated photon flux above 1\,TeV as a function of the time with respect to the periastron passage indicated with the vertical dashed line. The corresponding orbital phases (mean anomaly) are shown on the upper horizontal axis. The data from the 2004 (blue squares) \citep{psrb1259_hess05}, 2007 (red triangles) \citep{psrb1259_hess09}, and 2011 (green circles) observation campaigns are shown. For the 2004 and 2011 data the flux is shown in daily bins while for the 2007 data the flux is shown in monthly bins for clarity. The figure is taken from \citet{sushch_psrb}.}
  \label{overall_lc}
  \vspace{-10pt}
\end{figure}  

\begin{table*}[t]
  \small
  \centering
  \begin{threeparttable}[b]
    \caption{Analysis results of the H.E.S.S. data for the pre-flare and flare periods.}
    \label{analysis_res}
    \tabcolsep=0.11cm
    \begin{tabular}{c | c c | c c c }
      \hline
      \hline
      &&&&\\
      Dataset& Livetime [h] & Significance [$\sigma$]& $\Gamma$ & $N_{0}$ [$10^{-12}\,\flux$]& Flux($E>1$ TeV) \\  
      &&&&&[$10^{-12}$ \fluxunits]\\
      \hline
      &&&&&\\
      
      Pre-flare & 2.65& 7.4& $2.94\pm0.52_{\mathrm{stat}} \pm 0.20_{\mathrm{syst}}$& $2.15\pm0.56_{\mathrm{stat}}\pm 0.43_{\mathrm{syst}}$ & $1.11\pm0.29_{\mathrm{stat}} \pm 0.22_{\mathrm{syst}}$\\
      &&&&&\\
      Flare & 3.59& 8.5&  $3.26\pm0.49_{\mathrm{stat}} \pm 0.20_{\mathrm{syst}}$& $1.81\pm0.39_{\mathrm{stat}}\pm 0.36_{\mathrm{syst}}$& $0.80\pm0.22_{\mathrm{stat}} \pm 0.16_{\mathrm{syst}}$\\
      \hline
    \end{tabular}
    
  \end{threeparttable}
\end{table*}

In Fig. \ref{overall_lc}, the integrated photon flux above 1\,TeV 
as a function of time with respect to periastron (indicated by 
the dashed vertical line) is shown. The light curve compiles the data from all 
three periastron observation campaigns spanning from 100 days before to 
100 days after the periastron. The observed flux from the 
2010/2011 observation campaign (in green on Fig.\,\ref{overall_lc}) 
is compatible with the flux detected in 2004 at the similar orbital phases. 
%% Observation periods from 2004 and 2007 were separated 
%% in time with respect to the periastron position, i.e., observations in 2004 
%% were performed mainly after and in 2007 mainly before the periastron. 
%% Therefore, it was impossible to directly confirm the 
%% repetitive behaviour of the 
%% source by comparing observations of \psrb\ at the same orbital 
%% phases. In this perspective, 
Although the 2011 observations do not 
exactly overlap with the orbital phases of previous studies, 
they cover the gap in the 2004 data post-periastron light curve 
and the integrated flux follows the shape of the light curve, 
yielding a stronger evidence for the repetitive behaviour of the source. 
Moreover, the spectral shape of the VHE \gammaray\ emission from \psrb\ around 
the 2010/2011 periastron passage is similar to what was observed during 
previous periastron passages. The photon index of 
$2.92\pm0.25_{\mathrm{stat}}\pm 0.2_{\mathrm{syst}}$ inferred 
from the 2011 data is well compatible with previous results 
\citep{sushch_psrb}.
%% \subsection{Light curves}
%% \label{lightcurves}
%% To check for variability of the source a light curve was produced on 
%% a night-by-night basis assuming the photon spectral index obtained 
%% in the spectral fit (Fig. \ref{night_by_night}). The spectral index was 
%% fixed at the value obtained in the spectral analysis of the total data because 
%% of the low statistics for each individual night. 
%% The light curve is consistent with a constant 
%% resulting in a mean flux of $(0.77\pm0.13)\times10^{-12}$ \fluxunits\ 
%% (horizontal line in Fig. \ref{night_by_night}) with $\chi^{2}/
%% \mathrm{NDF}=6.35/4$ 
%% (corresponds to the probability of 0.17; NDF is the number of degrees 
%% of freedom), yielding no evidence for variability in the 
%% seven-nights observation period. 
%% For each individual night the 
%% source is detected at a statistical significance level $>3\sigma$ 
%% except for the last point, whose significance is only $1.5 \sigma$.
%% \begin{figure}
%% \centering
%%  \resizebox{\hsize}{!}{\includegraphics{lc_nightly_psrb1259_model.pdf}}
%% \caption[Lightcurve]{Integrated photon flux above 1 TeV for individual observation nights. The solid horizontal line indicates the fit of a constant to the distribution. The flare start date is indicated by the dashed vertical line.}
%% \label{night_by_night}
%% \end{figure} 
\subsection{Search for the GeV Flare Counterpart at VHEs}
The absence of the flux enhancement during the GeV flare at radio 
and X-ray wavebands indicates that the GeV flare may be created 
by physical processes different from those responsible 
for the emission at other wavelengths. If one assumes that 
HE and VHE emission are created according to the same scenario, i.e. the
same acceleration and radiation processes and sites, 
then a flux enhancement of the same magnitude as 
observed at HEs should be also seen 
at VHEs. The VHE post-periastron data 
obtained with H.E.S.S. around the 2004 periastron passage 
do not show any evidence of a flux outburst at orbital phases 
at which the GeV flare is observed. However, the H.E.S.S. observations around the 2004 
periastron passage do not comprise the orbital phase when the GeV flare starts. 
Moreover, to compare H.E.S.S. 2004 data with the GeV flare observed after the 2010 
periastron passage, one has to assume that the GeV 
flare is a periodic phenomenon, 
which may not be the case.

\begin{figure}
  \centering
  \includegraphics[width = \linewidth]{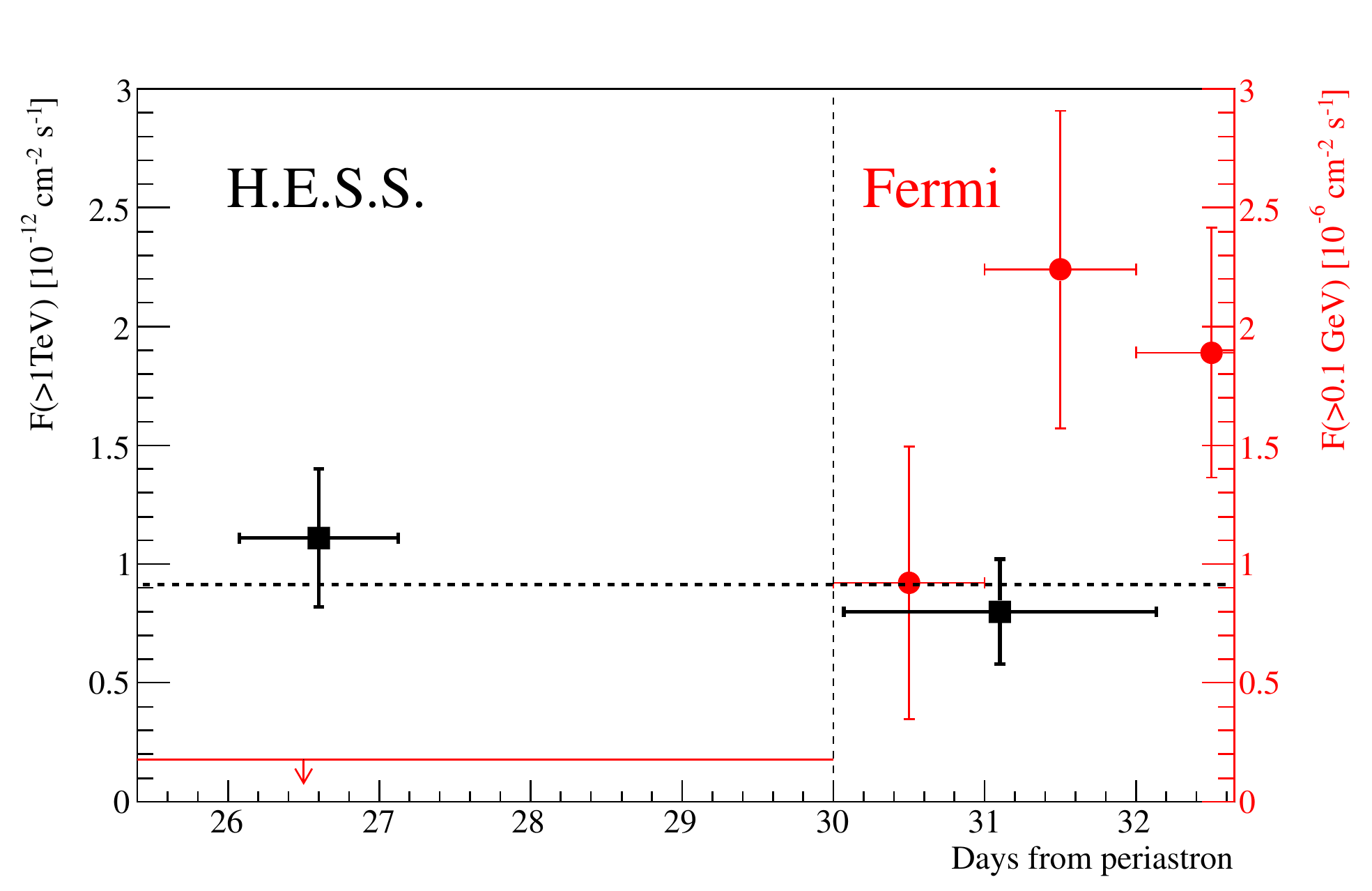}
   \vspace{-10pt}
  \caption{Integrated photon fluxes above 1 TeV for the pre-flare and flare periods (see text) are shown as black filled boxes. The dashed horizontal line shows the best fit with a constant. The HE data points above 0.1 GeV as reported by \citet{psrb1259_fermi} are shown as red filled circles. The flare start date is indicated by the dashed vertical line. The left axis indicates the units for the VHE flux and the right (red) axis denotes the units for the HE flux. The figure is taken from \citet{sushch_psrb}.
%% (Bottom) The spectral energy distribution of the HE-VHE emission. For the HE emission the overall flare spectrum is shown as reported by \citet{psrb1259_fermi}. Marking of the data points is the same as in the top panel. Solid lines denote the fit of the Fermi data only with the power law with exponential cut-off (red) and the fit of the \hess\ data only with the power law (black). The dashed black line denotes the fit of the Fermi (excluding upper limits) and H.E.S.S. data together with the power law.
}
  \label{pref_vs_flare_fig}
  \end{figure}  

The H.E.S.S. data taken between 9th and 16th of January 
in 2011 provide a three-day overlap 
in time with the GeV flare. Therefore, it is possible to directly 
study any flux enhancement in the VHE band on 
the time scale of the HE flare. To improve the sensitivity of the 
variability search the whole period of the H.E.S.S observations was 
divided into two time periods of almost equal length: 
"pre-flare" ($t_{\mathrm{p}}+26$\,d to 
$t_{\mathrm{p}}+29$\,d; before the HE flare) 
and "flare" ($t_{\mathrm{p}}+30$\,d to $t_{\mathrm{p}}+32$\,d; during the HE flare). 
These two datasets were analysed independently and revealed similar fluxes 
and significance levels (see Table \ref{analysis_res}). A spectral 
analysis of the two datasets shows that both spectra are consistent with a simple power law, yielding 
similar values of the spectral index (see Table \ref{analysis_res}). The two spectral indices are 
consistent with the one obtained for the total dataset \citep{sushch_psrb}.

A fit of the flux as a function of time with a constant showed no indication 
of the flux variability. The fit resulted in a mean flux of 
$(0.91\pm0.18)\times10^{-12}$ \fluxunits\ (black 
horizontal dashed line in Fig. \ref{pref_vs_flare_fig} top) and a 
$\chi^2$-to-NDF ratio of $0.73/1$, which corresponds to a $\chi^2$ 
probability of 0.39 \citep{sushch_psrb}.

To firmly reject the hypothesis of the GeV flare counterpart at VHEs, 
i.e. the flux enhancement of the same magnitude as GeV flare, 
the flare coefficient $\kappa$ was introduced as the ratio of 
the fluxes during the flare period and the pre-flare period 
\citep{sushch_psrb}. The ratio of the HE ($E > 0.1$ GeV) flux averaged over 
the three-day interval between $(t_{\mathrm{p}} + 30\,\mathrm{d})$ and 
$(t_{\mathrm{p}} + 32\,\mathrm{d})$ to the upper limit 
on the HE pre-flare emission (see Fig. \ref{pref_vs_flare_fig} (Top)) 
yields a lower limit on the HE emission flare 
coefficient $\kappa_{\mathrm{HE}} \geq 9.2$. The 
99.7 \% confidence level (equivalent of 3$\sigma$) 
upper limit on the VHE flare coefficient was estimated 
using the profile likelihood method (see \citet{sushch_psrb} 
for details). The obtained upper limit of $\kappa_{99.7\%}<3.5$ 
is lower than the observed lower limit on $\kappa_{\mathrm{HE}}$. This 
result suggests that the HE flare emission has a 
different nature than the VHE emission.
%% This conclusion is also supported by the inconsistency 
%% of HE and VHE emission spectra (see Fig. \ref{pref_vs_flare_fig} bottom). 
%% The joint fit of the Fermi and H.E.S.S. data points with the simple 
%% power law (the dashed line on  Fig. \ref{pref_vs_flare_fig} bottom) 
%% results in a fit probability of 0.004 and, hence, fails to explain 
%% ya the combined HE/VHE emission even ignoring the Fermi upper limits, which 
%% cannot be taken into account in the fit procedure. Moreover, the Fermi 
%% upper limits at $1-100$ GeV violate any reasonable model 
%% that would be able to explain the HE and 
%% VHE emission together. The Fermi spectrum alone is consistent with the 
%% the power law with exponential cut-off $E^2\mathrm{d}N/\mathrm{d}E = 
%% N_{0} \left( E/0.1 \mathrm{GeV} \right)^{-p} e^{-E/E_{\mathrm{cutoff}}}$ with 
%% the index $p = 0.16 \pm 0.32$, the cut-off energy of 
%% $E_{\mathrm{cutoff}} = 0.5\pm0.2$ GeV and the normalisation 
%% $N_{0} = (4\pm0.4)\times 10^{-10}$ erg cm$^{-2}$ s$^{-1}$. 
%% The fit probability is 0.27 \citep{sushch_psrb}.
\section{Modeling of the VHE Emission}
Several models have been proposed to explain the VHE emission 
from the source. In a hadronic scenario, the VHE \gammaray\ emission 
could be produced by the interaction 
of the ultrarelativistic pulsar wind particles with the dense equatorial disk 
outflow with subsequent production of $\pi^{0}$ pions and hence VHE 
$\gamma$-rays \citep{kawachi2004, neronov2007}. 
However, the detection of the source 
before the expected disk passage in 2007 casts doubts on the hadronic scenario, 
suggesting that the VHE emission should be created at least partly by 
leptonic processes \citep{psrb1259_hess09}. Within the leptonic scenario, 
VHE emission from \psrb\ is explained by the inverse Compton (IC) scattering of 
shock-accelerated electrons on stellar photons \citep{tavani_psrb1259, kirk99, 
dubus2006, khangulyan2007}.

In this paper for the modeling of the VHE emission the leptonic scenario 
is considered. Since the flaring HE emission was produced in a different 
scenario than the VHE emission, the HE flux measured by \emph{Fermi}-LAT 
can be used to constrain parameters which describe the electron population. 
Few possible explanations for the nature of the HE flare are discussed 
in the literature \citep{khangulyan2011_post, bogovalov_2008, dubus2010, 
kong2012}, but their discussion is beyond the scope of this paper. 

Since at VHE the system was 
observed for a short period and the lightcurve does not show any variability on 
this time scale, the time dependence can be neglected, and thus, one population 
of electrons can be assumed. The energy distribution of the electron 
density is assumed to follow a power-law with an exponential 
cut-off
\beq
\frac{dN_{\mathrm{e}}}{d\gamma} = K_{\mathrm{e}} \gamma^{-p} e^{-\frac{\gamma}{\gamma_{\mathrm{max}}}},
\enq
where $p$ is the electron spectral index, $K_{\mathrm{e}}$ is the normalisation parameter 
%, which physically corresponds to the number of electrons 
with $\gamma = 1$, and $E_\mathrm{max} = mc^{2}\gamma_{\mathrm{max}}$ is 
the electron cut-off energy. It is further assumed that 
this distribution of electrons 
already accounts for all kinds of energy losses and that 
these losses are not important on the considered 
time scale. The absorption due to the pair production of the individual 
$\gamma$-rays is not taken into account in the calculation of the 
resulting \gammaray\ flux, as it does not play an important role in the 
post-periastron phase \citep{dubus_2006}. 
The electrons are assumed to be distributed 
isotropically at the spherical termination shock. The stellar radiation 
at the location of the pulsar wind termination shock plays the role of 
target photons. The target photon field has then a Planckian distribution 
with a temperature $T_{\ast} = 30000 K$ and an energy density of 
1 erg\,cm$^{-3}$, which corresponds to the distance between the 
star and the shock at the observed orbital phase, assuming that 
the distance from the pulsar to the shock is much smaller than 
the distance from the star to the shock. Although the 
model is rather simple and does not account for 
a number of different processes which take place in the system in the 
vicinity of the emission region, it can still give an understanding 
of the population of electrons required to produce the observed VHE 
\gammaray\ flux.

 \begin{figure}[t]
    \centering
    \includegraphics[width=\linewidth]{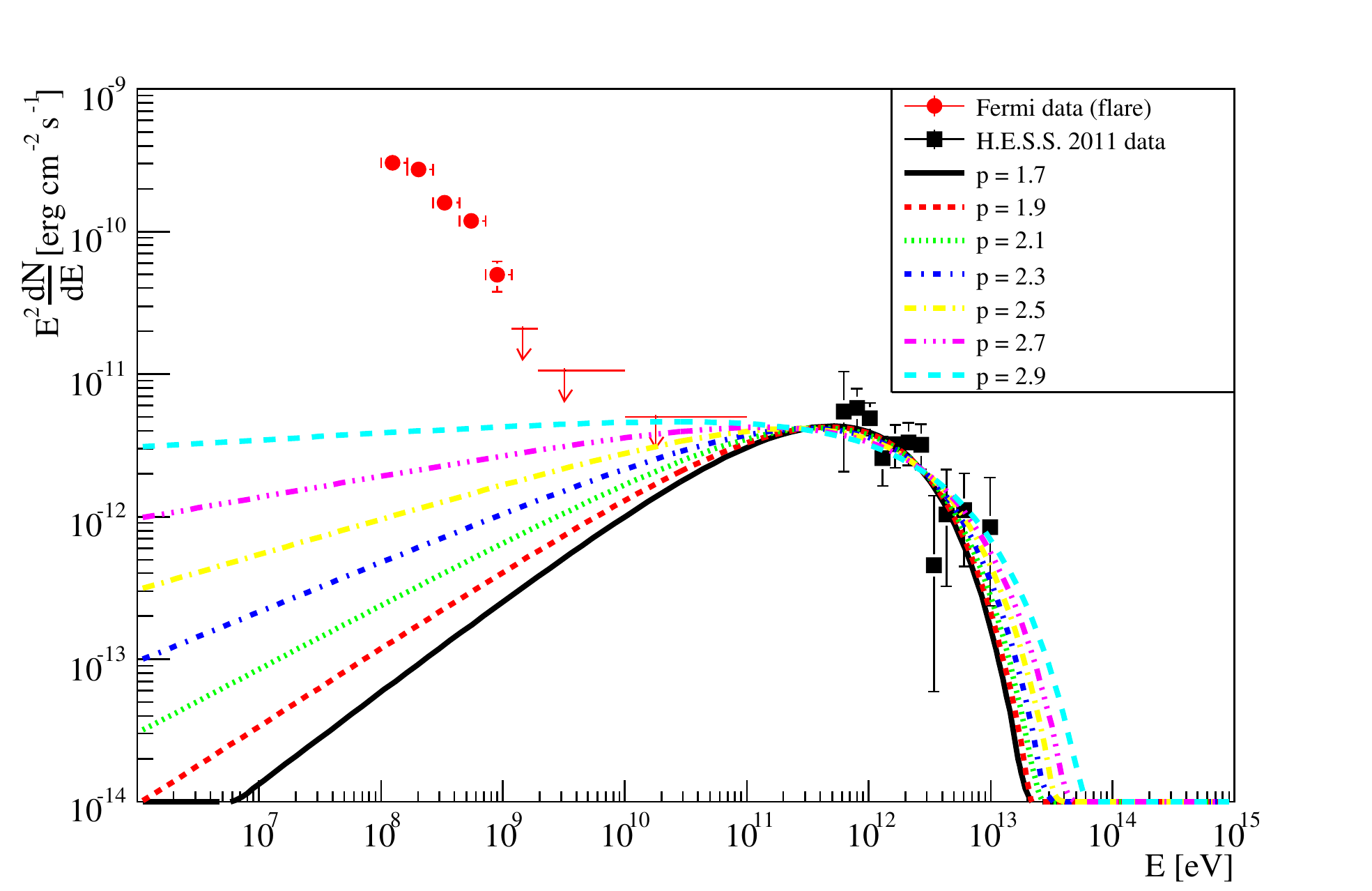}
     \vspace{-10pt}
    \caption[IC emission models]{Inverse Compton emission models for different spectral indices of the electron energy 
distribution. H.E.S.S. data are shown with black squares and \emph{Fermi}-LAT data are shown with red circles.}
  \label{SED_fermi_hess}
 \vspace{-10pt}
\end{figure}  
 
To constrain the IC flux the \emph{Fermi} upper limit value at 
100\,GeV is used. If one assumes that $E_{\mathrm{UL}} = 100$\,GeV is 
far from the cut-off energy, the IC emission at this energy would depend 
only on two parameters, $K_\mathrm{e}$ and $p$. Therefore, for each 
assumed $p$, one can calculate the upper limit on $K_\mathrm{e}$. Then, 
for each pair of $p$ and $K_{e}$ one can estimate $E_\mathrm{max}$ 
by fitting the IC emission to the H.E.S.S. data (Fig. \ref{SED_fermi_hess}). 
In Table \ref{Ke_vs_p_tab}, upper limits on $K_{\mathrm{e}}$ and corresponding 
$E_\mathrm{max}$ obtained from the fit of the H.E.S.S. data are shown 
for different values of the spectral index. The fit probability is also 
given in the table and is $\gtrsim0.5$ for all studied 
spectral indices. Based on 
the obtained parameters of the electron distribution 
an upper limit on the total energy in electrons $W_{\mathrm{tot}}$ 
which is needed to produce the VHE emission not exceeding 
the HE emission upper limit can be calculated. It is 
calculated assuming the minimum energy of the electron $E_{\mathrm{e, min}} = 1$ 
GeV. Depending on the assumed spectral index it 
varies from $4.0\times10^{43}$ erg for $p = 1.7$ to 
$1.7\times10^{46}$ erg for $p = 2.9$. Assuming that 
the spin-down luminosity of the pulsar, 
which is about $8\times 10^{35}$ erg/s, is transferred 
to the electron acceleration 
with the 100\% efficiency a lower limit on time 
required to accumulate $W_\mathrm{tot}$ can 
be derived. The lower limit on time ranges between 
1.5 y for $p=1.7$ to 674 y for $p=2.9$. For the 
range of spectral indices from 1.7 to 1.9 the lower limit on 
time is less than 3.4 y, i.e. the orbital period of the pulsar. 
This would mean that electrons accelerated at the shock during 
one orbital period may lose all their energy producing IC emission 
close to periastron and this process may repeat every 3.4 y. 
A possible extension of the electron spectrum down to energies lower 
than 1\,GeV would not change results significantly for $p<2.0$.

\begin{table}[ht]
\small
\centering
\vspace{-10pt}
\caption{Fit parameters for fixed values of the spectral index $p$}
\label{Ke_vs_p_tab}
\begin{tabular}{c c c c c }
\hline
\hline
%\\
$p$ & $\frac{K_\mathrm{e}}{(4 \pi D^{2})}$, cm$^{-2}$& $E_{\mathrm{max}}$, TeV& Fit prob.& $W_{\mathrm{tot}}$, erg\\
\hline
1.7& $2.2\times{10^{2}}$& 6.2& 0.72&$4.0\times10^{43}$\\ 
1.9  & $5.0\times{10^{3}}$& 7.1& 0.70&$6.9\times10^{43}$\\
2.1 & $1.1\times{10^{5}}$& 8.4& 0.68&$1.5\times10^{44}$\\
2.3 & $2.6\times{10^{6}}$ & 10.1& 0.65&$4.0\times10^{44}$\\
2.5 & $5.8\times{10^{7}}$ & 12.8& 0.61&$1.3\times10^{45}$\\
2.7 & $1.3\times{10^{9}}$ & 17.3& 0.56&$4.5\times10^{45}$\\
2.9 & $2.9\times{10^{10}}$ & 26.1& 0.50&$1.7\times10^{46}$\\
\hline
\end{tabular}
\end{table}

\section{Summary}
The binary system \psrb\ was monitored by \hess\ around the periastron passage 
on 15th of December 2010. The observed flux and 
spectral shape agree well with what was measured during 
previous periastron passages. The observations were 
performed at similar orbital phases as around the 2004 periastron passage, 
strengthening the evidence for the repetitive behaviour of the source at VHEs.

H.E.S.S. observations were part of a joint MWL campaign that also included 
radio, optical, X-ray, and HE observations. A spectacular flare observed 
at HEs with Fermi LAT overlapped in time with the H.E.S.S. observations. 
A careful statistical study showed that the HE flare does not have a 
counterpart at VHEs, indicating that the HE and VHE emissions are produced 
in different physical scenarios. The measured HE flux is then used to 
constrain the emission produced through the inverse Compton scattering of 
shocked relativistic electrons of stellar photons which is responsible 
for the VHE flux. It is shown that for the range of assumed electron spectral 
indices from 1.7 to 1.9 the lower limit on 
time is less than the orbital period of the pulsar.

\bigskip % extra skip inserted
%% \begin{acknowledgments}
%% The authors wish to thank JACoW for their guidance in preparing
%% this template.

%% Work supported by Department of Energy contract DE-AC03-76SF00515.
%% \end{acknowledgments}

\bigskip % extra skip inserted
\bibliography{references}
% \begin{thebibliography}{9}   % Use for  1-9  references
%\begin{thebibliography}{99} % Use for 10-99 references

%\bibitem{accelconf-ref}
%http://www.cern.ch/accelconf

%% \bibitem[Other et al.(1996)]{exampl-ref}
%% A.N. Other, ``A Very Interesting Paper'', EPAC'96, Sitges, June
%% 1996.
  %% 
%\bibitem{templates-ref}
%http://www.cern.ch/accelconf/templates.html

%\end{thebibliography}

\end{document}